# Synthetic gauge fields and Weyl point in Time-Reversal Invariant Acoustic Systems

Meng Xiao*, Wen-Jie Chen*, Wen-Yu He, Z. Q. Zhang and C. T. Chan[+]

*Department of Physics and Institute for Advanced Study, Hong Kong University of Science and Technology, Clear Water Bay, Hong Kong, China*

* These authors contributed equally to this work.

[+] Correspondence address: phchan@ust.hk

**Abstract**

Inspired by the discovery of quantum hall effect[1-3] and topological insulator[4-8], topological properties of classical waves start to draw worldwide attention[9-24]. Topological non-trivial bands characterized by non-zero Chern numbers are realized with external magnetic field induced time reversal symmetry breaking [10, 11, 23, 25] or dynamic modulation[12, 13]. Due to the absence of Faraday-like effect, the breaking of time reversal symmetry in an acoustic system is commonly realized with moving background fluids[19, 26], and hence drastically increases the engineering complexity. Here we show that we can realize effective inversion symmetry breaking and effective gauge field in a reduced two-dimensional system by structurally engineering interlayer couplings, achieving an acoustic analog of the topological Haldane model[3, 27]. We then find and demonstrate unidirectional backscattering immune edge states. We show that the synthetic gauge field is closely related to the Weyl points[28-30] in the three-dimensional band structure.

**Tight-Binding Model**

To illustrate how the idea works, we start with a simple tight binding model. The lattice structure is shown in Fig.1a, which is an AA stacked honeycomb lattice comprising a honeycomb lattice in the xy plane and periodic along the z direction. We introduce a nearest neighbor tight-binding model to describe this system. Its acoustic analogue will be discussed later in the following text. The Hamiltonian $H$ of this system consists of two parts: The intralayer part $H_0$ and interlayer part $H_1$. $H_0$ describes the coupling among atoms on the same layer, and it can be written as

$$H_0 = \sum_{i,k} \varepsilon \left( a_{i,k} a^\dagger_{i,k} - b_{i,k} b^\dagger_{i,k} \right) + \sum_{\langle a_{i,k}, b_{j,k} \rangle} \left( t_n a_{i,k} b^\dagger_{j,k} + \text{H.c.} \right), \quad (1)$$

where $a$ ($b$) and $a^\dagger$ ($b^\dagger$) are the annihilation and creation operators on sublattice site, respectively. Each lattice is specified by subscripts ($i$, $k$), where the first labels the position in each layer and the second labels the number of layer. The first term of $H_0$ represents the onsite energy difference between these two sublattices. The second term of $H_0$ represents the intralayer hopping between nearest sublattices. We assume the intralayer hopping $t_n$ to be real and constant. And hence when $\varepsilon = 0$, $H_0$ is identical to graphene described by nearest neighbor approximation. The first Brillouin Zone (BZ) of this AA stacked honeycomb lattice is shown in Fig. 1c, together with the definitions of some special points in the reciprocal space. As this system is periodic along the z direction, $k_z$ is a good quantum number. For each fixed $k_z$, and if we consider the dispersion and transport in the $xOy$ plane, the three-dimensional (3D) system can be reduced to an effective two-dimensional (2D) system with a unit cell as shown in Fig. 1b. The corresponding first BZ is in Fig. 1d, which represents a plane cut with the specified $k_z$ in original first BZ in Fig. 1c.

We can now introduce different kinds of interlayer coupling, among which, we choose two special examples (Figs. 1e and 1g, where hopping is non-zero only between connected sites) to illustrate our idea. We emphasize here that the system is time-reversal symmetric. In Fig. 1e, the hopping amplitudes are different for two different sublattices, and we use different colors, magenta and cyan, to illustrate this difference. In this case, the interlayer hopping part $H_1$ is given by

$$H_1 = \sum_{\langle i;k \rangle} \left( t_a a_{i,k} a^\dagger_{i,k+1} + t_b b_{i,k} b^\dagger_{i,k+1} + \text{H.c.} \right), \quad (2)$$

where $t_a$ and $t_b$ ($t_a \neq t_b$ and both are real) represent the hopping terms at different sublattices. The corresponding Bloch Hamiltonian $H(\boldsymbol{k})$, obtained by Fourier transform, is given by

$$H(\boldsymbol{k}) = \begin{pmatrix} \varepsilon + 2t_a \cos(k_z d_h) & t_n \beta \\ (t_n \beta)^* & -\varepsilon + 2t_b \cos(k_z d_h) \end{pmatrix}, \quad (3)$$

where $\beta = \exp(-ik_x a) + 2\cos(\sqrt{3}k_y a/2)\exp(ik_x a/2)$, and $a$ is the distance between two sublattices, $d_h$ is the interlayer distance and $\boldsymbol{k} = (k_x, k_y, k_z)$ is the Bloch wave vector. The eigenvalue of $H(\boldsymbol{k})$ are given by $E = (t_a + t_b)\cos(k_z d_h) \pm \sqrt{|t_n \beta|^2 + [\varepsilon + (t_a - t_b)\cos(k_z d_h)]^2}$. The first term under the square root is a function of $k_x$ and $k_y$, which vanishes along the KH direction. The last term under the square root shows the onsite energy difference and the interlayer coupling difference. If $[\varepsilon + (t_a - t_b)\cos(k_z d_h)]$ is nonzero, the inversion symmetry of the reduced 2D hexagonal lattice is broken, as illustrated pictorially in Fig. 1f. When $|t_a - t_b| > \varepsilon$, there exist special values $k_z = \pm\arccos[\varepsilon/(t_b - t_a)]/d_h$ where $H(x, y, k_z) = H(-x, -y, k_z)$. That implies that Dirac cones can form in the $k_x$-$k_y$ plane for these values of $k_z$. These special points correspond to Weyl point in the 3D band structure and we will discuss these Weyl points in detail later.

Another interesting example is shown in Fig. 1g, in which we consider a chiral kind of interlayer coupling. We will see that this is a realization of the topological Haldane model[3]. The coupling coefficients $t_c$ are taken to be identical and real. For a fixed $k_z$, the interlayer hopping becomes next nearest neighbor hopping in the reduced 2D system with a complex hopping coefficient $t_c e^{i\phi}$, where $\phi = \pm k_z d_h$ depends on whether the hopping is clockwise or anticlockwise as illustrated in Fig. 1h. In Fig. 1h, we use red arrows to represent the direction along which $\phi = k_z d_h$. After a hopping loop along the direction guided by the red arrows, the total phase accumulated is $3k_z d_h$. This means the flux enclosed by this loop is non-zero and $3k_z d_h$. Same calculations can also be performed along other loops inside the unit cell. In Fig. 1h, we use dotted and crossed black circle to represent the sign of local flux. While the total flux inside each unit cell is zero, we have non-zero local flux. The chiral interlayer coupling in the $z$-direction introduces Peierls phase for the hopping parameters in 2D for any non-zero $k_z$, which means that we can achieve a staggered synthetic gauge field in a system that is "static". This is different from the proven paradigm of using dynamical perturbation to induce synthetic gauge fields. This kind of chiral interlayer coupling is rare for electronic systems, but we will see that it can be easily implemented in classical waves using simple artificial structure. We note that the coupling as illustrated in Fig. 1e and Fig. 1g can be tuned independently, and their dependences on $k_z$ are also different. For a system with both the coupling shown in Figs. 1e and 1g, one can then tune the system across a topological transition point by changing the value of $(t_a - t_b)/t_c$ or $k_z$. The phase diagram is shown in the Appendix A.

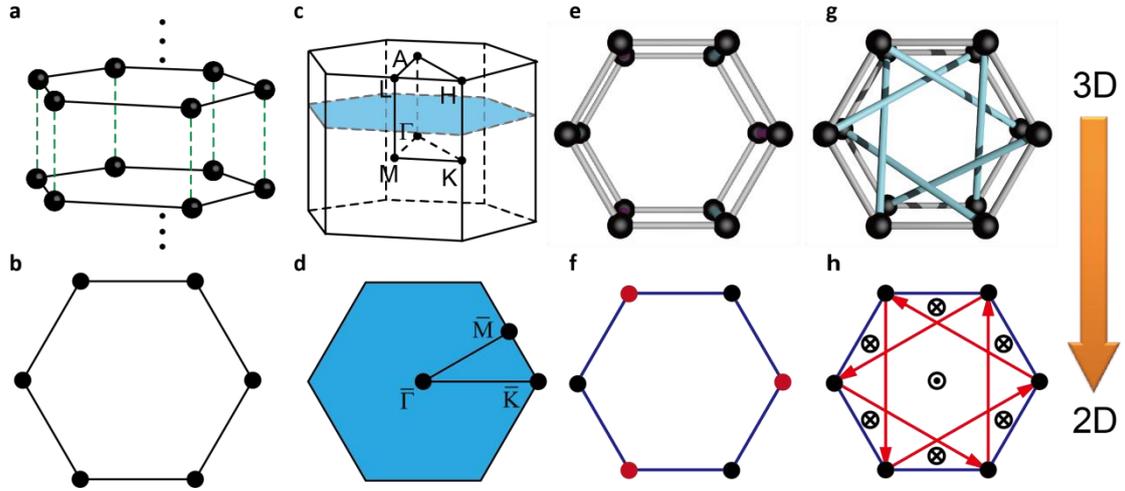

FIG. 1 **a, b**. AA stacking of a hexagonal lattice and its unit cell. **c**. The reciprocal space of the system shown in **a**. For each fixed $k_z$, we have a reduced first Brillouin Zone as shown in **d**. Different interlayer couplings introduce different effects in reduced two-dimensional lattice. **e, f**. Unequal interlayer coupling at different sublattice sites corresponds to broken inversion symmetry in 2D. **g, h**. Chiral interlayer coupling generates effective gauge field at each fixed $k_z$. The red arrows in **h** represent the positive phase hopping direction, and dotted and crossed black circles represent the direction of local flux. This is an analog of Haldane model.

**Breaking Inversion Symmetry**

After illustrating our idea with a tight binding model, we now show how to realize this idea in real acoustic structures. Let us consider a periodic array of acoustic cavities linked together by tubes as shown in Fig. 2a and 2b. The resonance cavities can be viewed as "meta-atoms" and the hopping strength between the meta-atoms can be tuned easily by changing the radius of the connecting tubes which is roughly proportional to the cross sectional area of the tubes when the cross sectional area is not too large. We found this simple acoustic implementation gives results that can be interpreted with the tight binding model. In Figs. 2a and 2b, we give the top and side views of the unit cell of our designed acoustic system. Light blue in Fig. 2b represents area where hard boundary condition is applied and the system is filled with Air (density $\rho = 1.3 \text{ kg/m}^3$, sound velocity $v = 343 \text{ m/s}$). The intralayer (parallel to the $x$O$y$ plane) coupling are set to be equal by choosing the same radius ($w_0$) for all the horizontal connection tubes. The interlayer coupling (along the $z$ direction) are different by choosing different radii ($w_1 \neq w_2$) at different sublattice sites. The discrete resonance of an isolated cavity is broadened into a band dispersion through coupling via the connecting tubes. Here, we consider a fundamental mode (not the lowest frequency mode) whose pressure has a sinusoidal function along the z direction and no variation in the horizontal plane. This mode has the symmetry of a $p_z$ orbital. In Fig. 2d, we show the band dispersions in the $k_x$-$k_y$ plane with different values of $k_z$. At $k_z = 0$, (red lines in Fig. 2d), the 2D dispersion in $k_x$-$k_y$ has a gap at the $\bar{K}$ point due to broken inversion system. As this reduced 2D system still has mirror symmetries with respect to the $x$O$z$ plane, the Berry curvature is an odd function in the reciprocal space and the two red bands are topologically trivial bands with zero Chern numbers. We found that at $k_z = 0.623$, there is a Dirac point at $\bar{K}$ as shown by the black lines in Fig.2d. This is consistent with the tight-binding model prediction that degeneracy for the reduced 2D system can be recovered at some finite $k_z$ value. The value of $k_z$, where the system has a Dirac cone is the $k_x$-$k_y$ plane, is slightly different from the tight binding prediction (Within tight binding model, the system has a Dirac cone at $k_z = 0.5$ with normalized unit $\pi/(h_c + l)$). This is because the introduction of the connection tubes changes the resonance frequencies of the cavities which is equivalent to $\varepsilon \neq 0$ in equation 3. In Fig. 2e, we show the band dispersion (black curves) along the KH direction (z-direction). The difference in freqeuency of the two bands along $k_z$ reflects the strength of inversion symmetry breaking as a function of $k_z$.

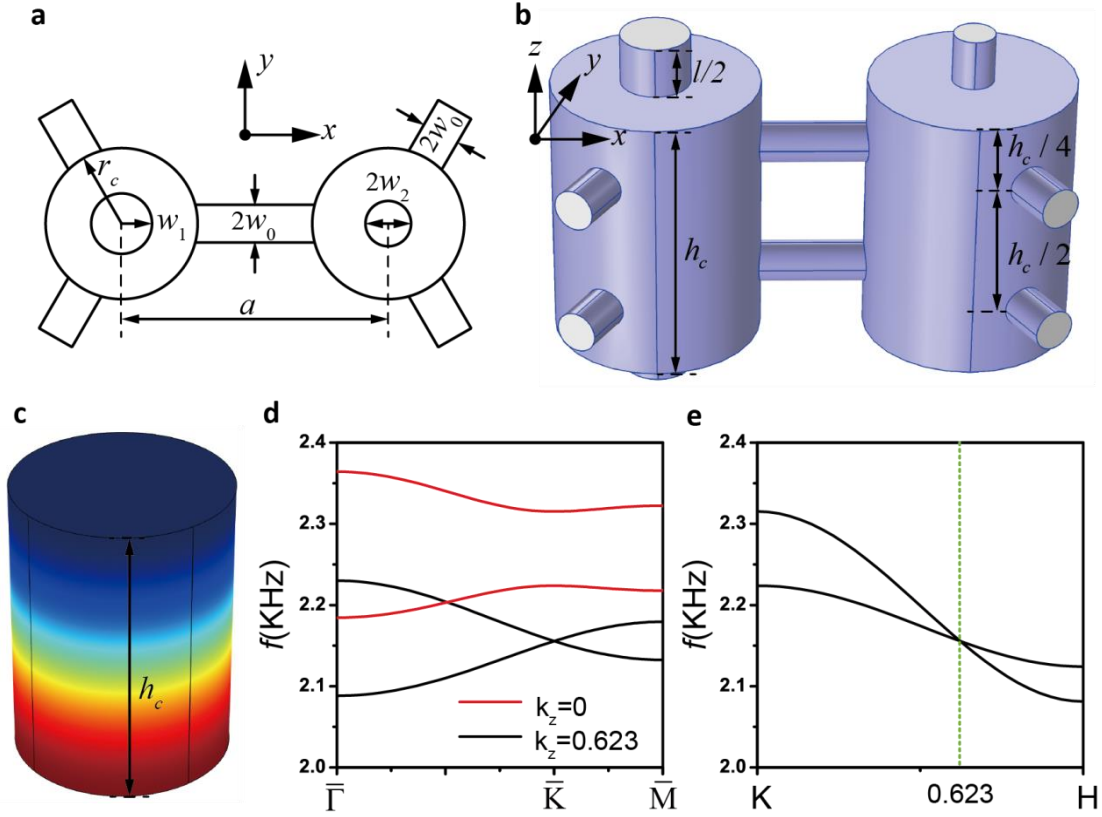

FIG. 2 **a**, **b**, Top view and side view of a unit cell of the acoustic system. **c**. The real part of the eigenpressure field of a cavity mode for a single resonator at 2144Hz, where red and blue represent positive and negative pressure, respectively. **d**. Band dispersion of the cavity mode in **c** for different $k_z$, where black curve has a Dirac point at $\bar{K}$ for a special value of $k_z$ and the degeneracy is lifted (red curve) otherwise. The Chern number of the two red bands are both C=0. **e**. Band dispersion along KH. The two bands crosses at a special point ($k_z$=0.623 with normalized unit $\pi/(h_c+l)$). This special point corresponds to a Weyl point of this acoustic system.

**Synthetic Gauge Field**

The realization of effective acoustic gauge fields using coupled acoustic resonators is conceptually straight forward. However, creating effective acoustic gauge fields with chiral coupling can go beyond the tight binding description. In Fig. 3, we design an acoustic system to show the effect of the synthetic gauge field. We emphasize that the hexagonal lattice symmetry in the *xOy* plane is not a necessary requirement to create the gauge field; but we choose to use the hexagonal system to demonstrate that we can achieve an acoustic analog of the Haldane model. Figs. 3a and 3b show the top view and side view of a unit cell. The chiral coupling is characterized by the relative rotation angle $\theta$ between the holes on the upper and lower boundaries of the planar (*xOy* plane) wave guides. When $\theta = 0$, the coupling is non-chiral and so the synthetic gauge field vanishes. The strength of gauge field depends on the rotational angle as well as the radius of the connection tube $r_1$. For details, please refer to the Appendix B.

We consider the lowest order acoustic mode and same as before, the wave guide is still filled with air. In Fig. 3c, Red/Black curves represent the band dispersion in the reduced 2D BZ at $k_z = 0 / k_z \neq 0$ representing the system with/without the effective gauge field. It is clear that the degeneracy at $\bar{K}$ is lifted by the gauge field. The effect of this synthetic gauge field can also be seen from the Chern number of each isolated band, and it is found to be $+1/-1$ for the lower/upper band when $k_z$ is positive. The strength of the gauge field can be seen from the width of gap at the $\bar{K}$ point. In Fig. 3d, we show the band dispersion along the KH direction. The width of gap reaches its maximum around the middle of the KH line.

Non-zero Chern number implies the existence of one-way edge mode in the boundary region between this system and a topological trivial system inside the common gap region. We construct a hexagonal ribbon with finite width along the *y* direction and periodic along the *x* direction (See Appendix C). We use hard boundaries to confine the sound wave in the *y* direction to be inside the acoustic system. The hard boundary condition can be regarded as a trivial band gap with zero decay length. Fig. 3e shows the projection band (gray) along the *x* direction and the dispersions of two surface states. Red and blue represent surface states localized at the lower and upper boundary in the *y* direction, respectively (See Appendix C). The two degeneracy points in Fig. 3d correspond to Weyl points[28-32] in the 3D band structure. If we consider a surface BZ spanned by $k_x$ and $k_z$, the allowed boundary modes for a given excitation frequency will trace out trajectories connecting two Weyl points that are the analogue of "Fermi Arcs" in electronic Weyl semi-metals[29]. (See Appendix D). If the width of the ribbon is large enough (larger than the decay length of the edge state), the edge state localized at one boundary cannot be scattered to the state localized at the other boundary. As there is no other state inside the complete gap region, the surface state cannot be scattered backward in the presence of impurity or external scatters as long as $k_z$ is still preserved. In Fig. 3f, we show the one-way property of the edge state. The edge state was excited on the left boundary (marked by the purple star) inside the gap region, and it propagates clockwise across corners and the defect without being backscattered. Black arrows are drawn to show the direction of the sound wave propagating. The direction of the sound wave propagation, clockwise or anticlockwise, depends on the sign of $k_z$.

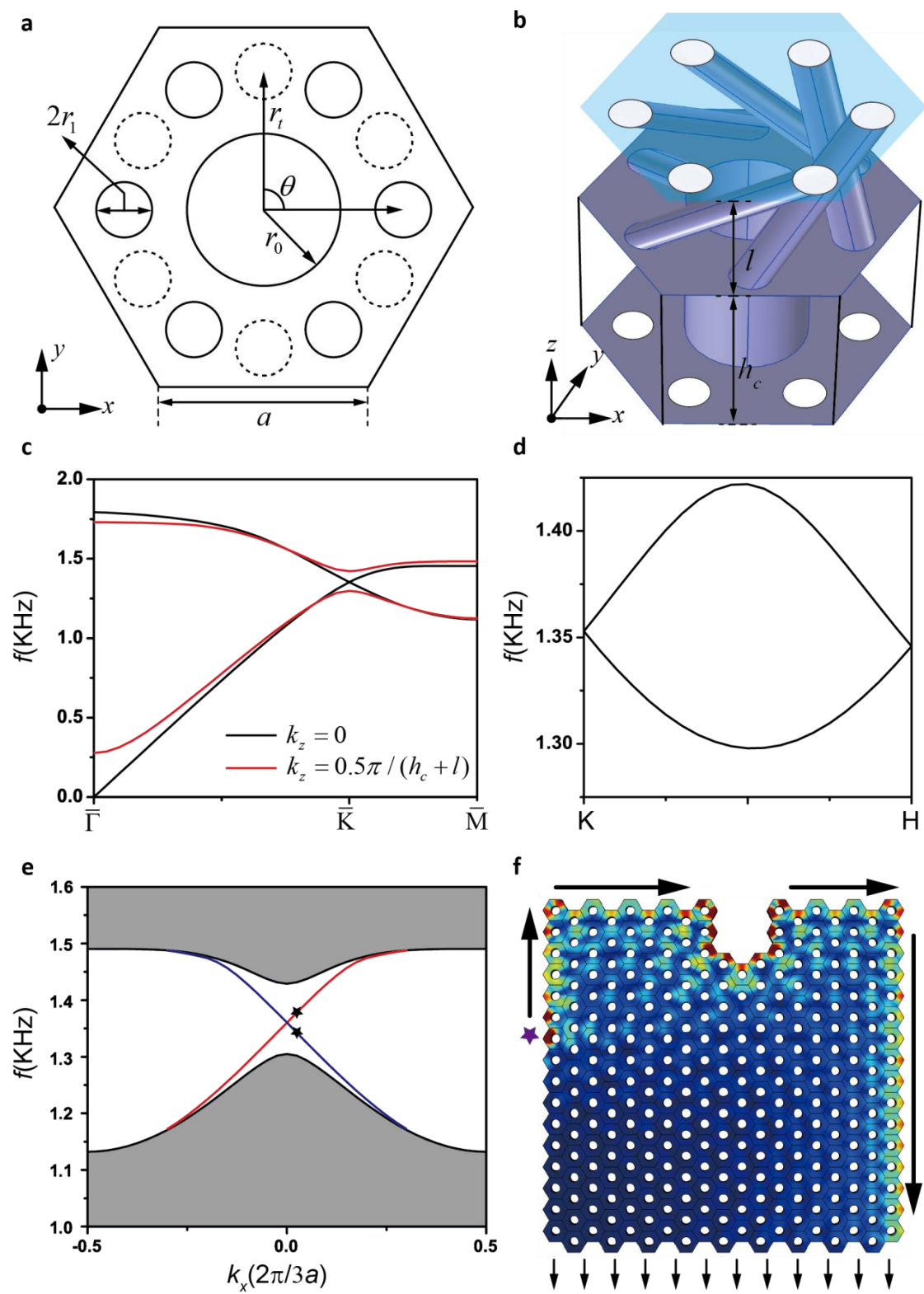

FIG. 3 **a**, **b**. Top view and side view of a unit cell of the acoustic system. The dashed circles and solid circles in **a** represent the holes opened at the upper side and lower side of the in-plane (*x*O*y*) sound wave guide, respectively. $\theta$ represents the rotation angle of the connection tubes. **c**. Band dispersion of the lowest mode for two values of $k_z$, where black/red represents bands without/with the effective gauge field. The Chern numbers of the two red band are C=+1 for lower band and C=-1 for upper band. The sign of Chern numbers can be changed by reversing the sign of $k_z$. **d**. Band dispersion along KH in the reciprocal space. K and H points are Weyl points of the 3D band structure. The strength of synthetic gauge field reaches its maximum at about $k_z = 0.5\pi/(h_c+l)$. **e**. Projection band (grey) along the *x* direction with $k_z = 0.5\pi/(h_c+l)$. Red and blue curve represent the surface states localized at the upper boundary and lower boundary of a ribbon, respectively. (See *Supplementary Material* Fig. 5, in which the eigenpressure fields of the states marked by the black stars are shown). **f**. Surface wave propagates anticlockwise across the corners and the defect of a finite system without being back scattered. Purple star marks the position of sound source and black arrows illustrate the direction of the sound wave propagation.

**Weyl Point**

We have demonstrated the existence of synthetic gauge fields in reduced 2D BZ. This synthetic gauge field is related to the Weyl points in the 3D band structure. In three dimensions, Weyl point dispersion is governed by Weyl Halmiltonian $H(\bm{k}) = \sum k_i v_{ij} \sigma_j$, $i, j \in \{x, y, z\}$ [28-32]. Weyl points are topological singular points in the reciprocal space with associated topological charges (or chirality, $c = \text{sgn}\left[\det(v_{ij})\right] = \pm 1$), which can be regarded as monopoles of Berry flux[30]. Berry flux emerges from the Weyl point with a positive charge and ends with the other Weyl point with a negative charge. A Weyl point is topologically stable which can only be eliminated by annihilating with another Weyl point with the opposite charge. The fact that Weyl points only exist along the high symmetry lines simplifies the analysis for finding Weyl points[28]. In Figs. 4a and 4b, we show the Weyl points in the reciprocal space of the two acoustic systems considered in Fig. 2 and Fig. 3, respectively. The system considered in Fig. 2 has mirror symmetry with respect to the *xOz* plane and this ensures that the two mirror symmetric Weyl points in the reciprocal space have opposite signs of charges, and hence the net charge on the horizontal light blue plane in Fig. 4a is zero. For any 2D band with an arbitrary (but fixed) value of $k_z$, the net Berry flux vanishes and the Chern number of this band is zero. In contrast, the system with chiral coupling [e.g. Fig. 3] does not possess mirror symmetry and remaining $C_6$ symmetry ensures all the Weyl points on the same $k_z$ plane have the same sign of charge. Meanwhile, the net charge of Weyl points inside the first BZ must vanish[33] and this means there must be at least two planes with different $k_z$ possessing Weyl points of different charges, e.g., the $k_z d_h / \pi = 0$ and $k_z d_h / \pi = \pm 1$ planes in Fig. 4(b) carry net topological charges of $+2$ and $-2$ respectively. Thus for arbitrary fixed $k_z$ lying between these two planes, the net Berry flux through the reduced 2D BZ is $2\pi$ and the Chern number is $\pm 1$ with sign decided by the sign of $k_z$[28, 29, 33]. The chiral coupling guarantees the non-zero Chern number which corroborates with the existence of synthetic gauge field.

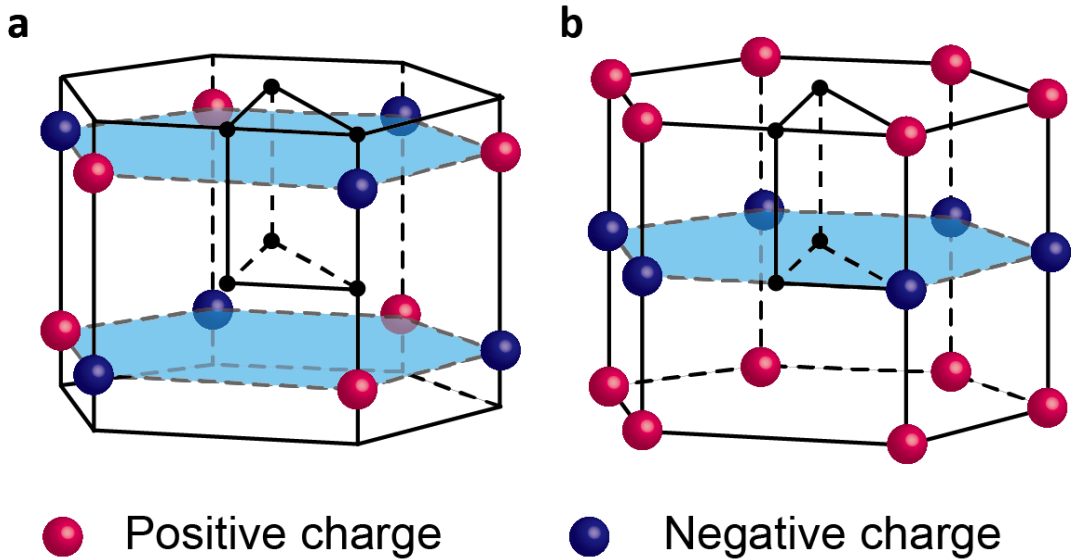

FIG. 4 (a) and (b) show the Weyl points for acoustic system studied in Fig. 2 and Fig. 3, respectively, where red and blue spheres represent Weyl points with positive and negative topological charges, respectively.

**Conclusion**

In summary, this study shows that we can realize topological acoustics in time-reversal symmetric systems with no time dependent perturbation and no moving parts. We can create synthetic gauge fields in static acoustic systems and such system is a realization of the topological Haldane model in acoustics. These acoustic systems also possess topological Weyl points and are thus realization of the Weyl Hamiltonian in sound waves. The effective gauge field induced by chiral coupling may stimulate new ideas of manipulating sound wave propagation, which have further implications in fields such as sound signal processing, sound energy harvesting, and noise protecting.

**Methods**

All simulations were performed using the commercial solver package COMSOL Multiphysics. Full three-dimensional geometry was used for all the simulations. The systems are filled with Air (Density $\rho = 1.3 \text{ kg/m}^3$, and the speed of sound $v = 343$ m/s). Eigenmode calculations were made to find the band structures as well as eigenmodes in Figs. 2c, 2d, 2f, 3c, 3d and 3e. Frequency calculation was made to find the one way transport behavior of the edge mode shown in Fig. 3f.

Figures. 2a and 2b, respectively, show the top view and side view of the unit cell studied in Fig. 2. The parameters used are given by $r_c = 3\text{cm}$, $a = 9\text{cm}$, $w_0 = w_2 = 0.6\text{cm}$, $w_1 = 1\text{cm}$, $h_c = 8\text{cm}$ and $l = 3\text{cm}$. Light blue in Fig. 2b represents area where hard boundary condition is applied, and white represents area where periodic boundary condition is applied with given Bloch wave vector. Figure. 2c shows the real part of pressure field of a cavity mode found at 2144Hz, where the parameters of the cavity are same as before, i.e., $r_c = 3\text{cm}$ and $h_c = 8\text{cm}$. Hard boundary condition is applied over all the cavity boundaries and the eigenmode calculation is then performed to find the eigen pressure distribution, where red/blue color indicates positive/negative local pressure.

Figures. 3a and 3b, respectively, show the top view and side view of the unit cell studied in Fig. 3. The parameters of the unit cell are given by $r_0 = 3\text{cm}$, $r_t = 5.5\text{cm}$, $a = 8\text{cm}$, $r_1 = 1.2\text{cm}$, $\theta = 90°$, $l = 5\text{cm}$, $h_c = 8\text{cm}$. The dashed circles and solid circles in Fig. 3a represent the holes opened at the upper side and lower side of the sound wave guide in the *xOy* plane, respectively. $\theta$ represents the rotation angle of the chiral kind connection tubes. In Fig. 3b, light blue represent area where hard boundary condition is applied, and solid white circles represent areas where floquet periodic boundary conditions are applied with a given Bloch wave vector. Floquet periodic boundary conditions are also applied on the side walls (For illustration, we set them to be transparent to expose the interior structure of the unit cell). To calculate the projection band shown in Fig. 3e, we construct a ribbon with finite length (16 unit cells) along the *y* direction. Floquet periodic boundary conditions are then applied for the side walls on the *x* direction and the *z* direction. The remaining boundaries are all set as hard boundaries. The Bloch wave vector along the *z* direction is fixed at $k_z = 0.5\pi/(l+h_c)$ in this calculation. In Fig. 3f, the purple star marks the position of our source. To couple wave inside our system, we use the Port boundary condition on one of the side boundary of a unit cell with plane wave excited at this port and working frequency at 1360Hz. All the side boundaries on the negative *y* direction are also set as Port boundaries to couple wave outside of our system, and we use black arrows to indicate the direction of wave propagation through these ports. All the boundaries on the upper side and lower side of the *z* direction are set as floquet periodic boundary condition with Bloch wave vector along the *z* direction given by $k_z = -0.5\pi/(l+h_c)$, where "+" or "−" sign of $k_z$ decides the direction (clockwise or anticlockwise) of surface wave propagation. All the remaining boundaries are set as hard boundary.

## Acknowledgments
The authors would like to thank Vic Law for discusssion. This work was supported by the Hong Kong Research Grants Council (Grant No. AoE/P-02/12).

**Acknowledgments**

The authors would like to thank Vic Law for discusssion. This work was supported by the Hong Kong Research Grants Council (Grant No. AoE/P-02/12).

# Appendix

## A. Phase diagram of the parameter space within tight binding model

In this section, we derive the phase diagram in the parameter space when both kinds of couplings (Figs. 1e and 1g) are included. The strength of the effective inversion symmetry breaking in reduced two-dimensional (2D) space is determined by the difference between the diagonal terms in equation (3) in the main text, i.e., $2\varepsilon + 2(t_a - t_b)\cos(k_z d_h)$. The strength of the synthetic gauge field is related to the local flux variation which is also a function of $k_z$. The competition between these two mechanisms determines whether the system is topologically trivial or nontrivial.

To derive the transition boundaries in the parameter space across which the system change from topologically trivial to topological nontrivial or vice versa, we can start from the distribution of the Weyl points. As we have said in the main text, Weyl point is a topological singular point with associated topological charge. Once we find the positions of all the Weyl points in the reciprocal space, we know the transition boundaries. As a start point, we set $\varepsilon = 0$ and the $\varepsilon \neq 0$ case can be obtained with a little more effort.

When both kinds of couplings as shown in Figs. 1e and 1g are considered, the corresponding Bloch Hamiltonian $H(\mathbf{k})$ becomes

$$H(\mathbf{k}) = \begin{pmatrix} 2t_a \cos(k_z d_h) + t_c f(k_z d_h) & t_n \beta \\ (t_n \beta)^* & 2t_b \cos(k_z d_h) + t_c f(-k_z d_h) \end{pmatrix}, \quad (A1)$$

where $f(k_z d_h) = 2\cos(\sqrt{3}k_y a - k_z d_h) + 4\cos(3k_x a/2)\cos(\sqrt{3}k_y a/2 + k_z d_h)$. $H(\mathbf{k})$ can be expanded in the form $H(k) = \sum_{i=0}^{3} c_i \sigma_i$, where $\sigma_i$ ($i \neq 0$) is the Pauli matrix and $\sigma_0$ is the identical matrix.

At a Weyl point, two bands become degeneracy, and this means all the coefficients $c_i$ ($i = 1, 2, 3$) must be zero simultaneously ($\sigma_0$ just contributes a global shift of the energy spectrum). The off diagonal terms only depend on $k_x$ and $k_y$, and $\beta$ vanishes at the vertical direction of the $K$ ($K'$) point, or in other words, the $\bar{K}$ ($\bar{K}'$) point in the reduced 2D BZ. And hence Weyl points can only exist along these high symmetry lines. All the Weyl points at the $\bar{K}$ ($\bar{K}'$) point have the same $k_z$ and the same sign of charge, we only need to analyze the Weyl points at one $\bar{K}$ point and one $\bar{K}'$ point.

At $\bar{K}$ point ($k_x = 0, k_y = \frac{2\pi}{a}\frac{2\sqrt{3}}{9}$), $f(k_z d_h)$ is simplified to be $f(k_z d_h) = 6\cos\left(\frac{2\pi}{3} + k_z d_h\right)$. Set $c_3 = 0$, we have

$$\frac{t_a - t_b}{3t_c} = \frac{\cos(2\pi/3 - k_z d_h) - \cos(2\pi/3 + k_z d_h)}{\cos(k_z d_h)}. \quad \text{(A2)}$$

We note that this curve diverges at $k_z d_h = \pm \pi/2$ where the effective inversion symmetry is preserved. At $\bar{K}'$ point ($k_x = 0, k_y = -\frac{2\pi}{a}\frac{2\sqrt{3}}{9}$), also set $c_3 = 0$, we have

$$\frac{t_a - t_b}{3t_c} = \frac{\cos(2\pi/3 + k_z d_h) - \cos(2\pi/3 - k_z d_h)}{\cos(k_z d_h)}. \quad \text{(A3)}$$

The only difference between equations A2 and A3 is the sign. Black curves in Fig. 5 show the relations in equations S2 and S3 and these curves are mirror symmetric with respect to the horizontal axis.

To analyze the topological property of a 2D band with fixed $k_z$, let us start from the case with only chiral interlayer coupling and focus on the evolution of Weyl points when we gradually turn on $|t_a - t_b|$. When $t_a - t_b = 0$, the distribution of the Weyl points as well as their associated charges are shown in Fig. 4b. When we gradually increase $t_a - t_b$ while fix $t_c$, the two Weyl points at the vertical direction of the $K/K'$ points move upward/downward. In Fig. 6, we show the distribution of Weyl points when $(t_a - t_b)/3t_c$ is non-zero and finite. We see from this figure that the Weyl points with positive and negative topological charges are located at different $k_z$ planes. Chern numbers of 2D bands with fixed $k_z$ can be obtained from counting the number of Weyl points[28, 29,] [33] inside the first BZ, and the results are shown on the right hand side of Fig. 6. Similar argument also works when $(t_a-t_b)/3t_c$ is negative, and hence we know the topological properties of all the regions separated by equations S2 and S3 in the parameter space and the results are shown in Fig. 5, where magenta, cyan and gray represent region with Chern number $+1$, $-1$ and 0 for the 2D band below the Weyl-point frequency, respectively.

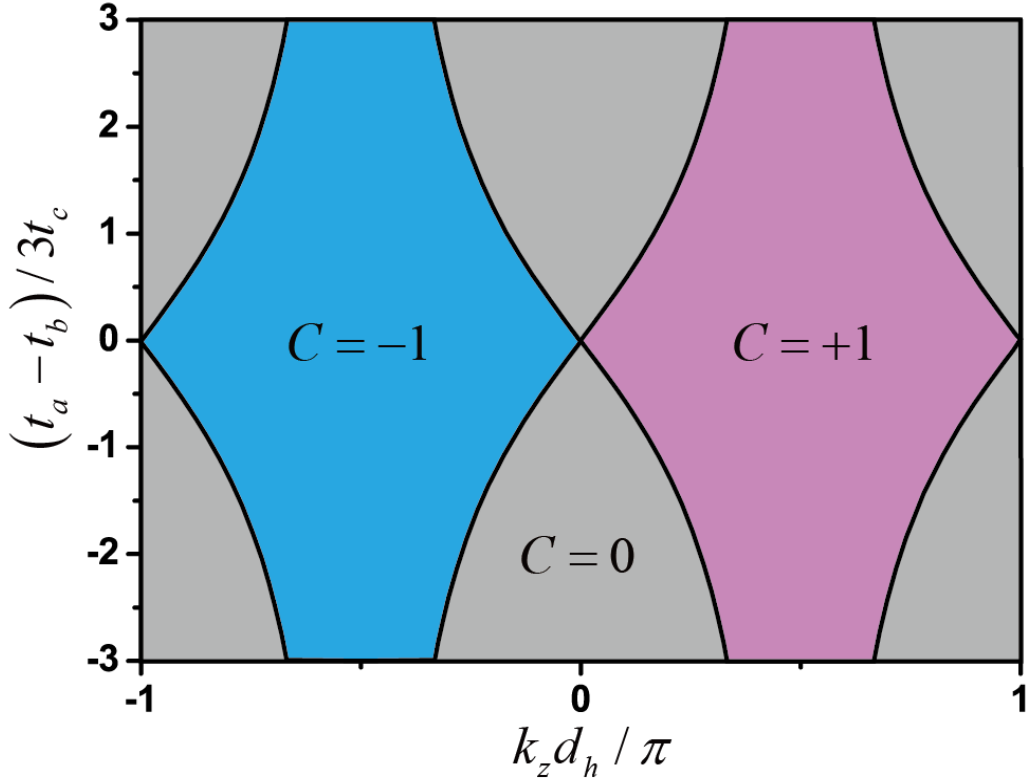

FIG. 5 | Phase diagram of the parameter space within tight binding model. The black curves give the boundary between the topologically trivial and non-trivial regions diverge at $k_z d_h/\pi = \pm 0.5$. Magenta, cyan and gray represent regions with Chern number $+1$, $-1$ and 0 for the band below the Weyl-point frequency, respectively.

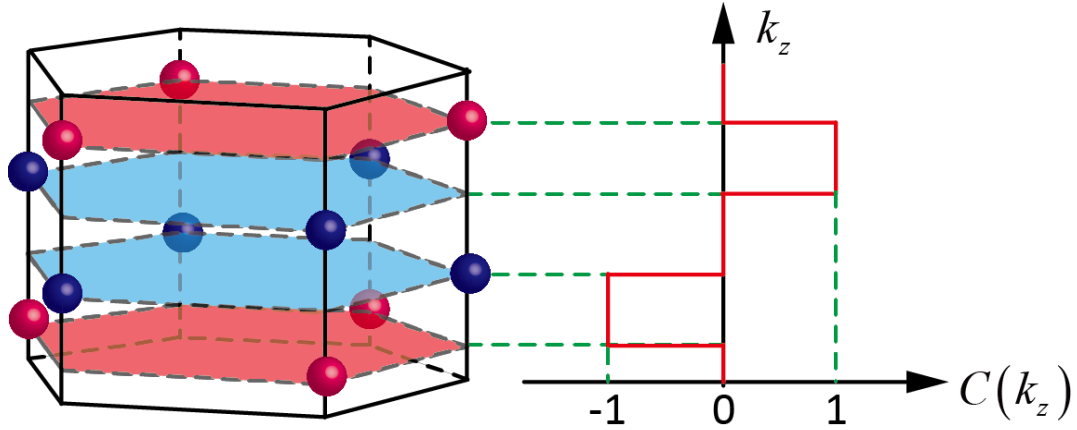

FIG. 6 Red solid line on the right hand side shows the value of Chern number of the 2D band below the Weyl-point frequency with a fixed $k_z$.

## B. Parameter dependence of the band gap at the middle of the KH line

For the chiral acoustic system shown in Fig. 3, the band gap at the $\bar{K}$ ($\bar{K}'$) point in the reduced 2D BZ is related to the strength of the synthetic gauge field. When there is no synthetic gauge field, e.g., black curve in Fig. 3c, the gap closes. As the width of the band gap depends on $k_z$ and it reaches its maximum around the middle of the KH line, we choose the width of band gap at the middle of the KH line as a reference quantity and study its dependences on the system parameters. Here we only focus on two parameters: The rotation angle $\theta$ and radius $r_1$. The definition of these two parameters can be found in Fig. 3. When $\theta=0$, the interlayer coupling is non-chiral and the two bands in Fig. 3d are degenerate, so the band gap closes. $r_1$ determines the interlayer coupling strength which in turn also determines the strength of the synthetic gauge field introduced by the chiral interlayer coupling. When $r_1$ is not too large, the coupling strength is proportional to the areas of the connection tubes. In Fig. 7, we show the band gap width at the middle of the KH line as functions of the rotation angle $\theta$ and radius $r_1$. Except for the rotation angle in Fig. 7a and $r_1$ in Fig. 7b, all the other parameters of the unit cell are same as those used in Figs. 3c and 3d. There are other parameters that also have some influence on the synthetic gauge field, e.g., the positions of the interlayer connection tubes. As the acoustic field is not homogenous inside the waveguide, the coupling strength also changes when the positions of the tubes change. The inhomogeneous of the acoustic field can also be seen from the ripple (it will be more obvious if we plot the first derivative.) in Fig. 7a.

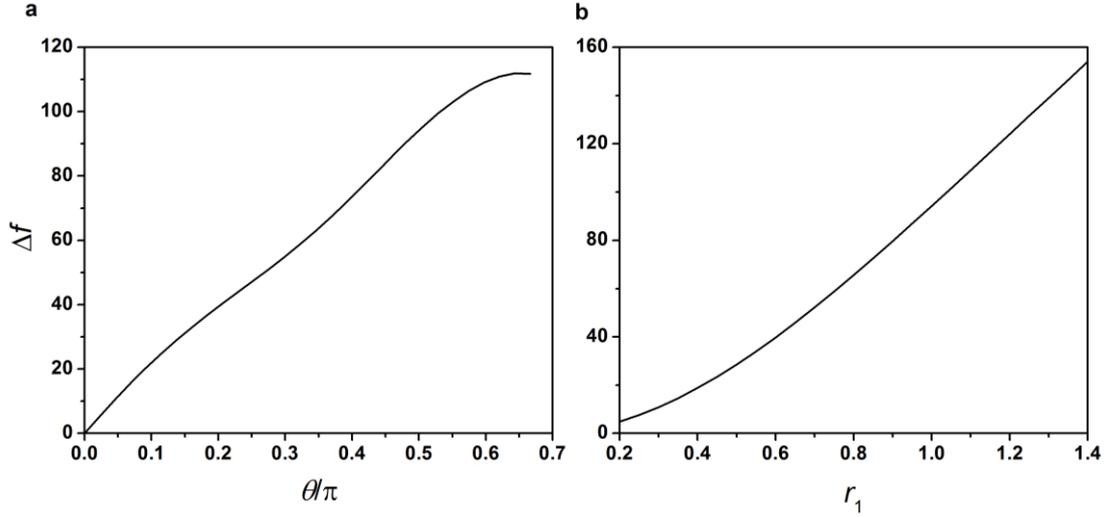

FIG. 7 The gap at middle of the KH line in the reciprocal space as a function of rotation angle $\theta$ (**a**) and radius $r_1$ (**b**) of the chiral connection tubes. Except for the rotation angle in **a** and $r_1$ in **b**, all the other parameters of the unit cell are the same as those used in Figs. 3c and 3d.

### C. Eigen pressure fields of the edge modes of a ribbon

In the main text we discussed the synthetic gauge field induced by the chiral interlayer coupling. When $k_z d_h \neq 0$ and $k_z d_h \neq \pi$, the Chern number of the two bands considered are non-zero. If we construct an interface between this system and a topological trivial system (such as a hard boundary), we can find one-way edge modes. In Fig. 8, we consider a ribbon with a finite length along the $y$ direction (confined by a hard boundary) and periodic along the $x$ direction. Periodic boundary condition is applied along the $z$ direction with $k_z(l+h_c)/\pi = 0.5$. With these settings, we obtain the projection band and two one-way edge modes localized at opposite edges. In Fig. 8, we show the eigen pressure fields of the edge modes found at $k_x = \pi/60a$, 1342Hz in Fig. 8a and 1378Hz in Fig. 8b, respectively. Here Red/blue color indicates positive/negative local pressure. The eigenfrequencies of these two modes are labeled by black stars in Fig. 3e.

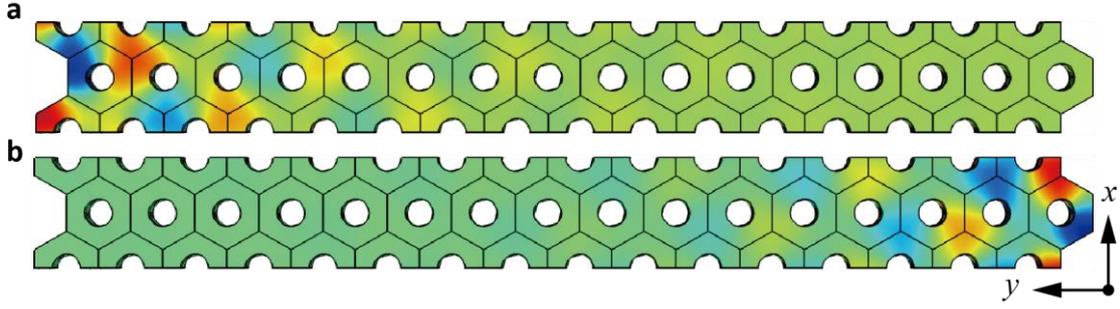

FIG. 8 The real part of the eigen pressure fields distribution of the edge mode found at $k_x = \pi/60a$, 1342Hz **a** and 1378Hz **b**. Red/blue color indicates positive/negative local pressure. The eigenfrequencies of these modes are labeled by black stars in Fig. 3e. Periodic boundary condition is applied along the z direction with $k_z(l+h_c)/\pi = 0.5$

### D. Surface state dispersion and the acoustic analogue of "Fermi Arcs"

If we consider a surface BZ spanned by $k_x$ and $k_z$, the surface wave dispersions (red and blue curves) in Fig. 3e become two curved surfaces as shown in Fig. 9, where red and blue represent the surface states localized at the upper and lower edges (y direction) of the ribbon considered in Fig. 8, respectively. The parameters used in this calculating are same as those used in Fig. 8. The two degeneracy points in Fig. 3d correspond to Weyl points[28, 29, 33] in the 3D band structure. In this surface BZ, the two Weyl points are located at $k_x = 0$, $k_z = 0$ and $k_z = \pi/d_h$ and the signs of the topological charges are opposite. Let us choose a working frequency of 1345Hz inside the band gap of the 2D projected band with a fixe $k_z$. This frequency coincides with the frequency of one of the Weyl points. We trace the trajectory of the edge modes of this frequency in the $k_x - k_y$ plane, shown as the black curve in Fig. 9. This trajectory is analogous to the "Fermi Arc" in Weyl semi-metals[29]. There are however differences. There is no "Fermi energy" for sound waves and hence we can only choose a working frequency which is that of the detector. We also note that the frequencies of the two Weyl points do not have to be at the same frequency. With the parameters we choose, the frequency difference of the two Weyl points is a few HZ, and this difference can be removed if we choose to tune the rotational angle $\theta$.

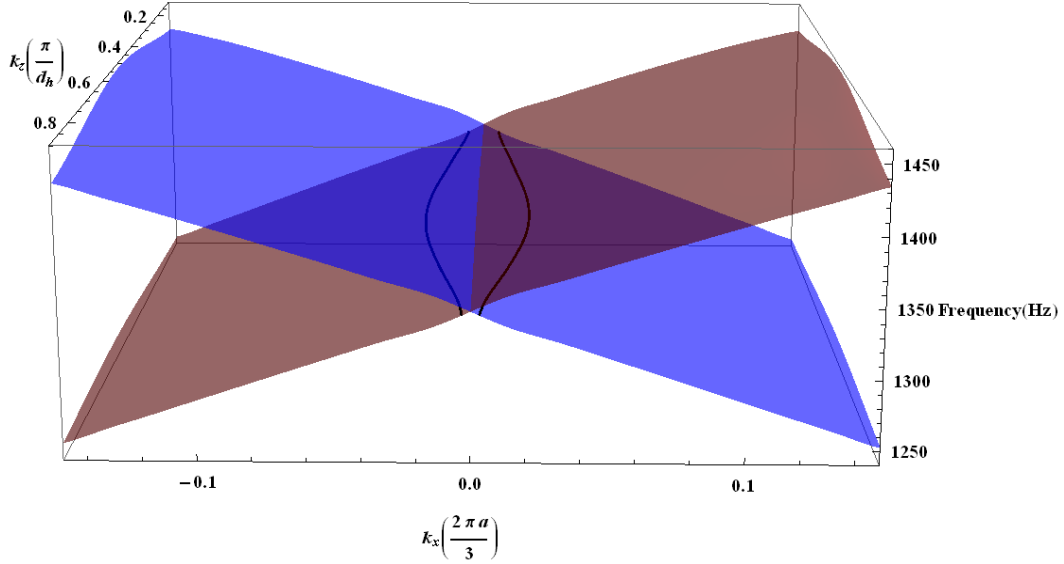

FIG. 9 Red and Blue represent the dispersions of surface states localized at the upper and lower edges (*y* direction) of the ribbon considered in Fig. 8, respectively. Black curves represent the trajectories of allowed edge modes at a working frequency of 1345Hz connecting two Weyl points with opposite topological charges. Such trajectories are analogues of "Fermi Arcs" in electronic Weyl semi-metals.